\documentclass[epsfig ]{emulateapj}

\def\micron{$\mu\mbox{m}$\,}
\def\gtorder{\mathrel{\raise.3ex\hbox{$>$}\mkern-14mu
             \lower0.6ex\hbox{$\sim$}}}

\keywords{quasars: general; quasars: emission line}
\slugcomment{}
\shorttitle{Discovery of three $z>5$ quasars in AGES }
\shortauthors{Cool et al.}

\begin{document}
\title{The Discovery of Three New $z>5$ Quasars in the AGN and
Galaxy Evolution
Survey }
\author{Richard J. Cool\altaffilmark{1},
Christopher S. Kochanek\altaffilmark{2},
Daniel J. Eisenstein\altaffilmark{1},
Daniel Stern\altaffilmark{3},
Kate Brand\altaffilmark{4},
Michael J. I. Brown\altaffilmark{5},
Arjun Dey\altaffilmark{4},
Peter R. Eisenhardt\altaffilmark{3},
Xiaohui Fan\altaffilmark{1},
Anthony H. Gonzalez\altaffilmark{6},
Richard F. Green\altaffilmark{7,4},
Buell T. Jannuzi\altaffilmark{4}, 
Eric H. McKenzie\altaffilmark{6},
George H. Rieke\altaffilmark{1},
Marcia Rieke\altaffilmark{1},  
Baruch T. Soifer\altaffilmark{8},  
Hyron Spinrad \altaffilmark{9},
Richard J. Elston\altaffilmark{6,10} }

\altaffiltext{1}{Steward Observatory, 933 N Cherry Avenue, Tucson,
AZ 85721;rcool@as.arizona.edu}
\altaffiltext{2}{Department of Astronomy, The Ohio State University,
140 West 18th Avenue, Columbus OH 43210}
\altaffiltext{3}{Jet Propulsion Laboratory, California Institute
of Technology, 4800 Oak Grove Drive, Pasadena, CA 91109}
\altaffiltext{4}{National Optical Astronomy Observatory, Tucson,
AZ 85726-6732}
\altaffiltext{5}{Princeton University Observatory, Peyton Hall,
Princeton, NJ 08544}
\altaffiltext{6}{Department of Astronomy, University of Florida,
Gainesville, FL 32611}
\altaffiltext{7}{University of Arizona, Large Binocular Telescope
Observatory, Tucson, AZ 85721}
\altaffiltext{8}{Division of Physics, Math, \& Astronomy and 
{\it Spitzer} Science Center, California Institute
of Technology, MC 314-6, Pasadena, CA, 91125}
\altaffiltext{9} {Department of Astronomy, University of California at Berkeley, Berkeley CA 94720}
\altaffiltext{10}{Deceased}

\begin{abstract}
We present the discovery of three $z>5$ quasars in the AGN
and Galaxy Evolution Survey (AGES) spectroscopic observations of the NOAO
Deep Wide-Field Survey (NDWFS) Bo\"{o}tes Field.  These quasars were selected
as part of a larger {\it Spitzer} mid-infrared quasar sample with no selection
based on optical colors.   The highest redshift
object, NDWFS J142516.3+325409, $z=5.85$, is the lowest-luminosity
$z>5.8$ quasar currently known.
We compare mid-infrared techniques for identifying $z>5$
quasars to more traditional optical techniques
and show that mid-infrared colors allow for selection of high-redshift
quasars even at redshifts where quasars lie near the optical stellar locus and
at $z>7$ where optical selection is impossible.   
Using the superb multi-wavelength
coverage available in the NDWFS Bo\"{o}tes field, we construct the
spectral energy distributions (SEDs) of high-redshift quasars from observed $B_W$-band to 24\micron
(rest-frame 600 $\mbox{\AA}$ - 3.7\micron). We show that the three
high-redshift quasars have
quite similar SEDs, and the rest-frame composite SED of low-redshift quasars from the literature 
shows little evolution compared to our high-redshift objects.
We compare the
number of $z>5$ quasars we have discovered to the expected number from
published quasar luminosity functions. While analyses of 
the quasar luminosity function are tenuous based on only three 
objects, we find that a relatively
steep luminosity function with $\Psi \propto L^{-3.2}$ provides the best 
agreement with the number of high-redshift quasars discovered in our survey.
\end{abstract}

\section{Introduction}

Understanding the evolution of quasars through
cosmic time allows us to study the history of accretion of matter
onto supermassive black holes in the nuclei of galaxies.
In addition, quasars at high redshift provide
the lighthouses needed to probe the conditions under
which galaxies initially formed near the epoch of re-ionization.
Samples of quasars at high redshift, however, are still quite small
and most high-redshift quasar surveys are only sensitive to the most
luminous examples of these objects.  We know very little
about the population of low-luminosity quasars at these early times.  

Surveys for low-redshift quasars have been quite successful and have led
to quasar samples suitable for detailed statistical
studies of the evolution of the quasar luminosity 
function to moderate redshifts
\citep{pgbqs,boyle2000,fan2001IV,fan2001III,wolf2003,croom2004,richards2005,barger2005,
brown2006,jiang2005,richards2006}.
Censuses  of high-redshift objects have opened the Universe even 
back to the era of reionization, but the number of such objects is still small.
Deep spectroscopic follow-up of $i$-dropout objects in the Sloan 
Digital Sky Survey (SDSS) \citep{york2000}
have found 19  quasars at $z>5.7$ \citep{xfanI,xfanII,xfanIII,xfanIV}.  \citet{zheng2000} and 
\citet{chiu2005} found 7 quasars at $z\gtorder5$ based on SDSS photometry 
which complement the 17 quasars at $5<z<5.4$ found in SDSS DR3 \citep{a2004b} spectroscopy
selected as outliers from the stellar locus \citep{schneider2005}.
SDSS imaging, however, is relatively shallow ($z_{AB}<20.5 \,\mbox{mag}$)
and thus the quasars found using SDSS photometry probe only the 
luminous tail of the quasar luminosity function.  Furthermore, the use of optical
photometry for selecting high-redshift quasars is biased against heavily 
reddened objects, as the observed optical flux from objects at $z>3$ probes the rest-frame ultraviolet 
emission which is most heavily affected by dust and the strong absorption from neutral hydrogen
in the intergalactic medium.

The difficulty of conducting deep, wide-area multi-color surveys has meant
that only a handful of lower-luminosity sources have been reported.
 \citet{stern2000} §found a faint
quasar ($M_B=-22.7$) at $z=5.50$ in a small area imaged approximately
4 magnitudes deeper than SDSS.  A quasar at $z=5.189$ ($M_B\sim-23.2$)
was selected based on 1 Ms of X-ray imaging in the {\it Chandra}
Deep Field North
\citep{barger2002}.  \citet{djorgovski2003} imaged the field around
the quasar SDSS J0338+0021 at $z=5.02$ and discovered a second
quasar at $z=4.96$,
a magnitude fainter than the SDSS detection limit ($M_B=-25.2$).
\citet{mahabal2005} imaged the field
around the $z=6.42$ quasar SDSS J1148+5251 and report the discovery
of a
faint ($M_B=-24.3$) $z=5.70$ quasar.

Unfortunately, the number of successful
searches for faint high-redshift quasars is comparable to the number of
surveys with negative results.  \citet{barger2003} found no
additional
high-redshift quasars in the full 2 Ms imaging of the {\it Chandra}
Deep
Field North that were not detected in 1 Ms of observation and
\citet{cristiani2004} found no high-redshift objects based on
deep X-ray imaging of the {\it Hubble} Deep Field North and {\it Chandra}
Deep Field
South.
\citet{willott2005} imaged 3.8 deg$^2$ of sky in
$i$ and $z$   and \citet{sharp2004}
imaged 1.8 deg$^2$ in $VIZ$ reaching 3 and 2 magnitudes deeper
than SDSS,
respectively, but neither survey detected any new low-luminosity,
high-redshift quasars.

In this paper, we present photometry and spectroscopy of three new
high-redshift ($z>5$) quasars using the multi-wavelength photometry
available in  the NOAO Deep Wide-Field Survey  
\citep[NDWFS;][Dey et al., {\it in prep.}]{jannuzidey1999} Bo\"{o}tes field and
spectroscopic observations from the AGN and Galaxy Evolution Survey 
(AGES; Kochanek et al., {\it in prep}),
which provides redshifts for several highly-complete samples of 
quasars to $I_{\hbox{\scriptsize AB}}=22$ mag, nearly 2 mag 
fainter than SDSS selects
high-redshift objects.
With our three newly discovered $z>5$ quasars,
we compare the
number density inferred from this survey to predictions based on
quasar luminosity functions in the literature. 

This paper is organized as follows:
in \S2, we summarize all the  multi-wavelength photometry used in the
paper and in \S3 we discuss our spectroscopic observations.  
We compare the mid-infrared selection of quasars utilized
in this work to optical criteria used in the past in \S4.
Finally, we place our survey in context with past studies of the 
quasar luminosity function and consider future high-redshift quasar searches in \S5.
All optical photometry presented here are corrected for
foreground galactic reddening using the dust maps of \citet{sfd1998}. 
We use AB magnitudes for all bands \citep{oke1974}, although the photometric catalogs
from the NDWFS and FLAMINGOS Extragalactic Survey \citep[FLAMEX;][]{elston2005}
present Vega magnitudes\footnote[1]{$B_{W,AB}=B_{W,{\rm Vega}}$, 
$R_{AB}=R_{{\rm Vega}}+0.20$, 
$I_{AB}=I_{{\rm Vega}}+0.45$, 
$J_{AB}=J_{{\rm Vega}}+0.91$, and $K_{s,AB} = Ks_{,{\rm Vega}}+1.85$}.  
Flux measurements from {\it Spitzer} are converted to AB magnitudes using 
$m_{AB} = 23.9-2.5 \,\mbox{log} (f_\nu / 1 \mu \mbox{Jy})$.
Also, when quoting optical, near-infrared, and IRAC photometry, we use
SExtractor \citep{sextractor} MAG\_AUTO magnitudes \citep[which are comparable
to Kron total magnitudes;][]{kron1980} 
due to their small systematic errors and uncertainties at faint fluxes.
When calculating luminosities, we use a $(\Omega_m, \Omega_\Lambda) =
(0.3, 0.7)$ flat cosmology and $H_0 = 70$ km s$^{-1}$ Mpc$^{-1}$.

\section{Multi-wavelength Photometry}

\subsection{NOAO Deep Wide-Field Survey}
We utilize the deep optical ($B_WRI$) and 
near-infrared ($K_s$) imaging of the 9.3 deg$^2$ 
Bo\"{o}tes field provided by the 
third data release from the NOAO Deep Wide-Field 
Survey \citep{jannuzidey1999}.  A full description of the
observing strategy and data reduction will be presented elsewhere
(Jannuzi et al., {\it in prep}; Dey et al., {\it in prep}) and the data can be obtained publicly
from the
NOAO Science Archive\footnote[2]{http://www.archive.noao.edu/ndwfs \\ http://www.noao.edu/noao/noaodeep}. 
The NDWFS catalogs reach $B_{W,\mbox{\scriptsize AB}}\sim26.5$, 
$R_{\mbox{\scriptsize AB}}\sim25.5$, $I_{\mbox{\scriptsize AB}}
\sim25.3$, and $K_{s,\mbox{\scriptsize AB}} \sim23.2$ at 50\% completeness.

\subsection{$z$Bo\"{o}tes}
We imaged 8.5 $\mbox{deg}^2$ of the sky, covering 7.7 $\mbox{deg}^2$
of the NDWFS Bo\"{o}tes
field, between 29 January 2005 and 31 March 2005 in the $z'$-band
with 90Prime \citep{williams2004} on the Bok 2.3m telescope on
Kitt Peak.
The 90Prime imager offers a 1 $\mbox{deg}^2$ field of view
when mounted at prime focus on the Bok telescope. Exposure
times range from 1--2.5 hours throughout the field with typical seeing of 1\farcs8. Images were
flatfielded using observations of the twilight sky.   For each night of observations, 
we stack all of the dithered $z'$-band images to produce a high signal-to-noise ratio image of the fringing
pattern in the detector.  This master fringe image is then scaled and subtracted from each 
object frame to remove the strong fringing pattern. 

We calibrate the astrometry and photometry of these $z'$-band images
using
public imaging from the Sloan Digital Sky Survey DR4 \citep{york2000,a2005}.
Cross comparisons  between $z$Bo\"{o}tes and SDSS show a 0\farcs1 rms
dispersion in the astrometry and a 5\% scatter in the photometry for bright stars ($z<19$).
This 5\% scatter is likely a combination
of photometric calibration errors in SDSS \citep[expected to be on the order to
2\%;][]{ivezic2004} and imperfect fringe removal and flat
fielding in the zBo\"{o}tes imaging.
Furthermore, we find that objects observed in overlapping $z$Bo\"{o}tes
fields have an rms scatter of 5\% in the final photometry.  The astrometry
between the NDWFS and zBo\"{o}tes are slightly offset ($\Delta \alpha\approx0\farcs4, \Delta \delta \approx0\farcs1$).  When matching objects in each zBo\"{o}tes field to 
the NDWFS catalogs, we remove the local mean astrometric offsets in both right ascension and 
declination from zBo\"{o}tes positions, resulting in astrometry that agrees to 0\farcs2 rms.     
As the
exposure times and observing conditions were variable throughout
the survey field, the limiting depth of the catalog depends on
location in the survey area.  The typical $3\sigma$ depth is 22.5 mag for
point sources in a 5 arcsecond diameter aperture.  Full details of the data reduction and a
full release of the $z'$-band imaging catalogs will be presented in a future data-release 
paper (Cool, {\it in prep}).

\subsection{FLAMEX}
The FLAMINGOS Extragalactic Survey (FLAMEX) \citep{elston2005}
provided some of the $J$ and $K_s$ photometry used in this work. The
Florida Multi-object Imaging Near-IR Grism Observational Spectrometer
(FLAMINGOS) on the Kitt Peak 2.1m telescope was used to image 4.7
$\mbox{deg}^2$
of the NDWFS Bo\"{o}tes field to a limiting depth of $K_{s,AB} \approx
21.1$ detecting approximately 150,000 sources ($5\sigma$). The FLAMEX 
catalogs are publicly available\footnote[3]{http://flamingos.astro.ufl.edu/extragalactic/overview.html}.

\pagebreak
\subsection{IRAC Shallow Survey}
The IRAC Shallow Survey \citep{eisenhardt2004} observed 8.5
$\mbox{deg}^2$
of the sky in the NDWFS Bo\"{o}tes region at 3.6, 4.5, 5.8, and 8.0
\micron
with the IRAC instrument \citep{fazio2004} on {\it Spitzer}.
This survey
found $\approx$ 270,000, 200,000, 27,000, and 26,000 sources brighter
than $5\sigma$ limits of 12.3, 15.4, 76, and 76  $\mu$Jy
(corresponding to limits of 21.2, 20.9, 19.2, and 19.2 AB mag)
in each
of the four IRAC bands \citep{eisenhardt2004,stern2005}.

\subsection{MIPS Imaging}
The NDWFS Bo\"{o}tes field was also observed at 24, 70, and 160
\micron with the Multiband Imaging Photometer for {\it Spitzer}
\citep{rieke2004} as part of the {\it Spitzer} IRS team's Guaranteed
Time Observing programs \citep{houck2005}; only the 24 \micron
photometry
is considered here.  The 24 \micron imaging covers 8.22
$\mbox{deg}^2$ of the NDWFS Bo\"{o}tes field and reaches a 80\%
completeness
limit of 0.3 mJy \citep{brown2006}.
\begin{deluxetable*}{cccccccc}
\tablecolumns{8}
\tablenum{1}
\tablewidth{0pt}
\tabletypesize{\scriptsize}
\tablecaption{Optical and Near-Infrared Photometry}
\tablehead{
  \colhead{Object Name} &
  \colhead{Redshift} &
  \colhead{$B_W$} & 
\colhead{$R$} &
\colhead{$I$} &
\colhead{$z'$} &
\colhead{$J$} &
\colhead{$K_s$}}
\startdata
NDWFS J142937.9+330416 &  5.39 & $>27.1$ & $22.84\pm0.09$ & $21.24\pm0.05$ & $21.02\pm0.07$ & $20.40\pm0.12$ & $20.81\pm0.1
9$ \\
NDWFS J142729.7+352209 &  5.53 & $>26.3$ & $24.20\pm0.21$ & $21.85\pm0.07$ & $21.93\pm0.11$ & \nodata & $20.41\pm0.19$ \\ 
NDWFS J142516.3+325409 &  5.85 & $>25.8$ & $23.99\pm0.11$ & $21.57\pm0.06$ & $20.68\pm0.06$ & \nodata & \nodata 
\enddata
\tablenotetext{}{\scriptsize All magnitudes are on the AB system}
\end{deluxetable*}

\begin{deluxetable*}{ccccccc}
\tablecolumns{7}
\tablenum{2}
\tablewidth{0pt}
\tabletypesize{\scriptsize}
\tablecaption{{\it Spitzer} Photometry}
\tablehead{
  \colhead{Object Name} &
  \colhead{Redshift} &
  \colhead{$[3.6 \, \mu m]$} &
  \colhead{$[4.5 \, \mu m]$}   &
  \colhead{$[5.8 \, \mu m]$}   &
  \colhead{$[8.0 \, \mu m]$}   &
  \colhead{$[24 \, \mu m]$} }
\startdata
NDWFS J142937.9+330416 &  5.39 & $20.81\pm0.08$ & $20.24\pm0.09$ & $>19.2$ & $19.6
5\pm0.13$ & $17.31\pm0.15$ \\
NDWFS J142729.7+352209 &  5.53 & $20.27\pm0.06$ & $19.84\pm0.07$ & $>19.2$ & $19.8
8\pm0.14$ & $17.53\pm0.18$ \\ 
NDWFS J142516.3+325409 &  5.85 & $20.36\pm0.06$ & $19.91\pm0.07$ & $>19.2$ & $20.7
4\pm0.18$ & $>17.7$ 
\enddata
\tablenotetext{}{\scriptsize All magnitudes are on the AB system}
\end{deluxetable*}

\section{Spectroscopic Observations}
\subsection{AGN and Galaxy Evolution Survey}
The AGN and Galaxy Evolution Survey (AGES, Kochanek et al., {\it in
prep}) has obtained complete spectroscopic samples of galaxies
and quasars using several multi-wavelength selection techniques in
the NDWFS Bo\"{o}tes field.  Spectra of $\approx$ 20,000 objects were
taken with Hectospec, a 300 fiber robotic spectrograph on the
MMT 6.5m telescope \citep{fabricant1998,fabricant2005,roll1998}.
Data reduction was completed using HSRED, a modified version of the
SDSS spectroscopic pipeline.  Dome flat spectra were used to correct
for the high-frequency flat-field variations and fringing in the
CCD, and, when available, twilight sky spectra provided a low-frequency
flat-field
correction for each fiber.  Each Hectospec configuration has
approximately
30 fibers dedicated to measuring the sky spectrum which is
subtracted from each object spectrum.  Simultaneous observations
of F-type stars in each configuration are cross correlated against
a grid of Kurucz models \citep{kurucz} to derive a sensitivity
function for 
each observation, thus linking the observed counts to absolute
flux units.
The final reduced spectra cover the 3700$\mbox{\AA}$ to
9200$\mbox{\AA}$ spectral range at resolution $R\approx1000$.

AGES quasar target selection occurred in two stages.  In the first phase
of the project, AGES obtained redshifts of nearly
all point sources with $R_{AB}<21.7$ mag that were either X-ray sources with 4 or more counts in the 
the XBo\"{o}tes survey \citep{murray2005} or were MIPS 24\micron
sources brighter than 1mJy with
non-stellar 24\micron to $J$-band colors.   This led to a sample of
roughly 900 spectroscopically identified AGNs as well as redshifts
of nearly 9000 galaxies selected using a variety of other techniques.  
By combining the AGES galaxy and AGN redshift 
samples, \citet{stern2005} confirmed
the ease with which quasars could be recognized using mid-infrared
colors. In the second phase of AGES spectroscopy, we extended the optical flux limit to $I_{AB}=22$
and added IRAC color selection to our methods of identifying quasars.  
In detail, mid-infrared selected objects considered here
were required to have $[3.6]-[4.5]>-0.1$ and either $[3.6]<20.8$ or a
detection at 5.8 or 8.0\micron and be classified as point-sources in the NDWFS imaging.   
In AGES, targets were classified as point sources if their SExtractor CLASS\_STAR parameter was larger than 0.8 for at least one of the 
$B_W$, $R$, or $I$-bands.
The AGES mid-infrared selection cut used for point sources dispenses with
the $[5.8]-[8.0]$ color restriction given in \citet{stern2005} which
is intended to minimize the contribution of $z>1$ galaxies which
are not common to our optical flux limits.  While we only consider the sample of point-sources selected in the mid-infrared from AGES in this paper, it is worth noting that when selecting AGN from extended source in AGES, the vertical selection cuts
from \citet{stern2005} are imposed to reduce contamination of star-forming galaxies on the AGN sample.  The shift from
an $R$-band to an $I$-band spectroscopic limit led to a significant
increase in the number of $z>4$ quasars in the AGES database,
including three $z>5$ objects.   The second phase of AGES target selection also included a 24\micron-selected quasar
sample.  These objects were required to be point sources,  have $f_{24\mu m} > 0.3 \mbox{mJy}$,
 and have $I$-band 3\farcs0
diameter aperture magnitudes, $I_{3,AB}$, such that $I_{3,AB} > 18.45 - 2.5 \mbox{log} (f_{24\mu m}/\mbox{mJy})$. This $I$-band selection criterion rejects normal stars from the sample but retains AGNs.  Two of the three $z>5$ quasars selected based upon their IRAC colors were also selected based upon their 24\micron photometry.

\subsection{Auxiliary Observations}
In order to augment the AGES discovery spectra, 
spectroscopic observations of two of the high-redshift
objects discovered in this work were completed
with the Multi-Aperture Red Spectrograph (MARS) on the 4-m
telescope on
Kitt Peak and the DEIMOS spectrograph \citep{faber2003}
on the Keck-II telescope under poor observing conditions.  These data were reduced using standard
techniques.  Also, to achieve higher quality data, two of the objects described in this paper were observed multiple times within AGES itself.  Coaddition of all of the available 
spectra for each object was performed to improve the final signal-to-noise
ratio of the data.  

\begin{figure}
\centering{\includegraphics[angle=0, width=4in]{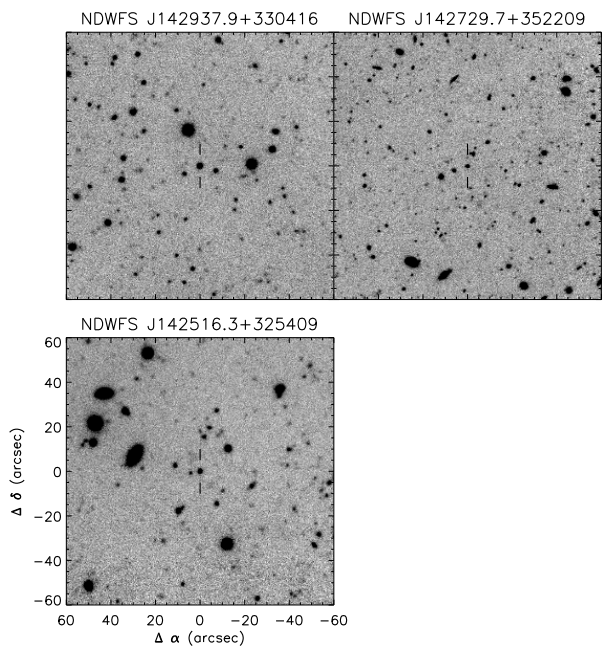}}
\caption{ \scriptsize Finding charts for the three $z>5$ quasars
discovered in the AGES survey.  The images are drawn from the
NDWFS $I$-band
imaging of the Bo\"{o}tes region and are each 2 arcminutes on a side.
North
is up and East is left.  The location of the quasar is marked
by vertical
lines in each panel.}
\label{fig:finder}
\end{figure}

\begin{figure}
\centering{\includegraphics[angle=0, width=3in]{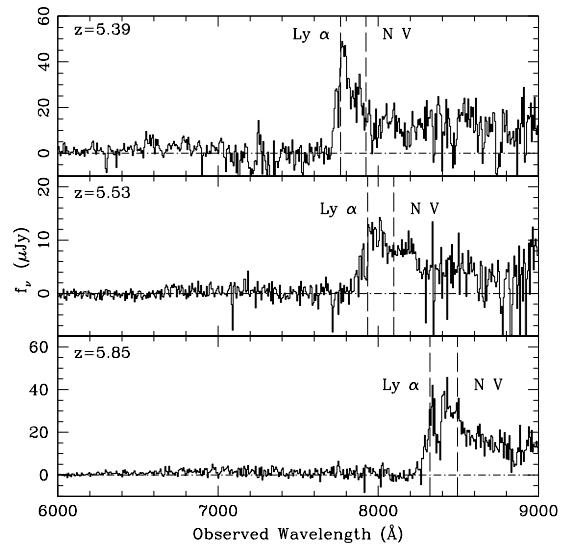}}
\caption{ \scriptsize Final stacked spectra for each of the
high-redshift quasars discovered by our work. These spectra
represent
the coadded spectra for the repeated observations with Hectospec as
well as observations with MARS and DEIMOS.  Two prominent quasar
emission lines, Ly$\alpha$ and NV, are marked with vertical
dashed lines and the measured redshift of each object is listed
in the upper left corner of each panel. All three of the quasars
show the strong Ly$\alpha$ emission line truncated
by neutral hydrogen absorption blueward of the emission line center
characteristic of high-redshift quasars.}
\label{fig:spectra}
\end{figure}

\section{Results}

We have identified three new high-redshift ($z>5$) quasars observed
as part of the $I_{\mbox{\scriptsize AB}}\leq22$ AGES spectroscopic sample. Figures
\ref{fig:finder} and \ref{fig:spectra} show the $I$-band finding chart and spectra for each of
the three high-redshift objects.  Tables 1 and 2 list the observed
photometry and redshifts of the three new 
high-redshift quasars and Table 3 shows
the derived rest-frame $B$-band luminosity and the luminosity of each object at
1450 $\mbox{\AA}$.  In calculating the rest-frame luminosities, we assume a power-law
SED of the form $f_\nu \propto \nu^{\alpha_\nu}$ where $\alpha_\nu=-0.5$.
The spectra shown in Figure \ref{fig:spectra} are the coadded spectra
from all the observations of each object:  NDWFS J142516.3+325409
was
observed three times in AGES as well as with MARS and DEIMOS,
NDWFS J142937.9+330416 was observed once in AGES and with MARS, and
 NDWFS J142729.7+352209 was observed twice within AGES.
Each of the discovered
objects show the clear signature of a broad asymmetric Ly$\alpha$
emission line.  As the signal-to-noise
ratios of the final spectra are only modest, redshifts for each object
were determined based upon the break in the Ly$\alpha$ line profile 
rather than on weak emission lines which are poorly detected.
 Quoted redshifts thus have errors of order $\Delta z \sim0.02.$

The quasars included in this paper were not selected with
standard dropout techniques which identify objects with very red
optical colors indicative of a strong spectral break.
These techniques have been successful in identifying quasars at
redshifts
above 5 \citep[e.g.][]{fan2000,xfanI,xfanII,xfanIII,xfanIV,
stern2000,djorgovski2003,mahabal2005}, but have several weaknesses.
At $z\gtorder5.5$, quasars cross the stellar locus in optical color-color space 
making optical selection problematic.  The inclusion of
near-infrared photometry can break this degeneracy to some extent,
but obtaining this data can be time consuming.

\begin{figure}[h!t]
\centering{\includegraphics[angle=0, width=3.5in]{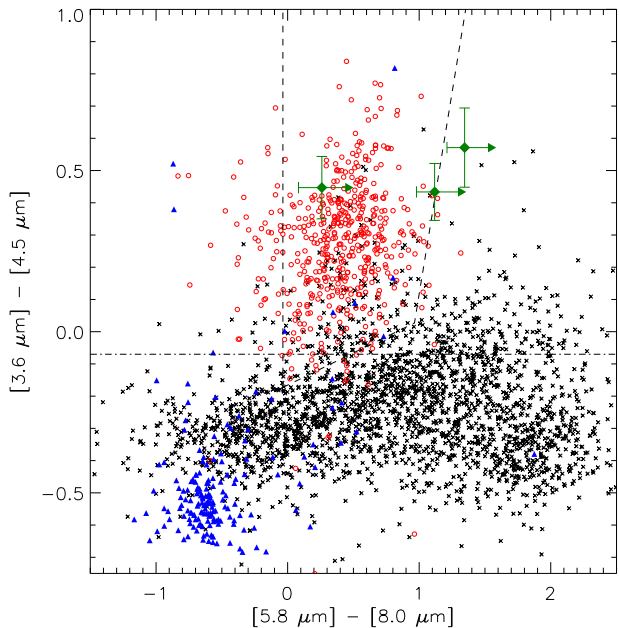}}
\caption{ \scriptsize 
 The mid-infrared colors of objects detected
in the AGES spectroscopic survey.  The symbol type and color
denote the redshift of each object. The x-marks show the
$0.01<z<0.5$ objects in AGES.  This sample is primarily composed
of galaxies,
although some low-redshift AGNs are also included.  The empty
circles (red) show
objects with $z>1.0$ and are dominated by AGNs.  The triangles (blue) mark the location of stars in this color-color space.
 The filled diamonds (green)
show the location
of the $z>5$ quasars in this color-color space.  The dashed
line shows the
region used by \citet{stern2005} to select AGNs.  The vertical
lines are used to
protect against the presence of $z>1$ galaxies in the AGN sample,
but as galaxies at $z>1$
are fainter than our optical flux limits, these cuts are not used
in the AGES mid-infrared
point-source quasar selection shown by the dot-dashed line.  When selecting AGN from extended 
objects in AGES, the vertical lines are used to reduce contamination from low-redshift star-forming galaxies.
}

\label{fig:irac}
\end{figure}

The high-redshift quasars studied here were all selected on the basis
of their mid-infrared colors using a scheme similar to
that presented in \citet{stern2005}. In brief, quasars, with their
roughly power-law
spectra, have redder mid-infrared colors than $z<1$ galaxies. 
The IRAC color-color space
for AGES objects shown in
Figure \ref{fig:irac}  illustrates
the separation of the high-redshift objects discovered here from
stars and
low-redshift galaxies.  We also indicate the criteria used for AGN selection
given by \citet{stern2005}  (dashed line) and the AGES point-source mid-infrared selection criterion (dot-dashed line).  
Extended sources are targeted as AGN in AGES if they meet the full \citet{stern2005} criteria.  Low-redshift ($z<0.5$) objects, which are predominately extended galaxies, are shown with x-marks while objects with $z>1$ are marked with circles and are dominated by unresolved AGN.  At $[5.8]-[8.0] \gtorder 1$, the locus of low-redshift galaxies
clearly crosses the AGES point-source mid-infrared selection criterion but remains outside the \citet{stern2005} selection region,\ illustrating the need for the separate point-source and extended-source selection criteria used by AGES.     The three high-redshift quasars studied here (diamonds with error-bars) are well-separated from low-redshift galaxies throughout this color-color space. 

\begin{figure}[ht]
\centering{\includegraphics[angle=0, width=3.5in]{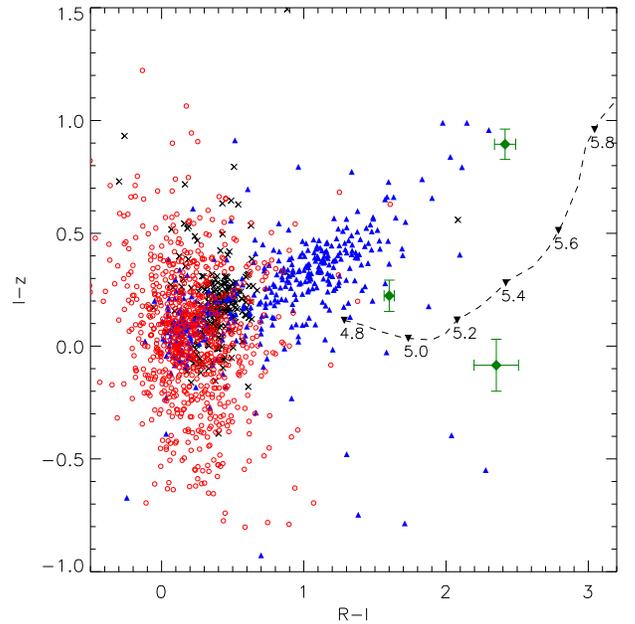}}
\caption{ \scriptsize  
Optical color-color space for AGES point
sources.
The symbols are defined as in Figure \ref{fig:irac}. Additionally, 
the dashed line shows the expected colors of quasars at $z>4.8$ predicted
using the SDSS quasar composite \citep{vandenburk2001} modified by the addition of
absorption blueward of Ly$\alpha$  using the prescription of \citet{songaila2002}.
 The three
high-redshift quasars discovered in this work occupy the same
location in
color-color space used in the past to optically-select $z>4.5$
quasars.  At
the same time, two of the three objects lie very near the stellar
locus
in this color-color space, illustrating one of the main difficulties
in using
optical colors alone to select high-redshift quasars.
}
\label{fig:riiz}
\end{figure}

One obvious question arises due to our mid-infrared selection of
high-redshift quasars: does the population of mid-infrared selected quasars differ from
quasars selected using more traditional optical techniques?  The spectra in
Figure \ref{fig:spectra}, though of only modest signal-to-noise
ratio, show
no obvious differences from objects presented in the literature
and would
suggest no strong difference between high-redshift quasars selected
based upon their optical or mid-infrared colors.  Figure \ref{fig:riiz}
shows the $I-z$
versus $R-I$ color-color space for point sources observed in AGES
as well as
the color track of  high-redshift quasars predicted
using the
\citet{vandenburk2001} SDSS quasar composite  modified
by the addition of 
absorption blueward of Ly$\alpha$ using the prescription of 
\citet{songaila2002}.
The three quasars discovered here reside in the region traditionally
used to
select quasars with $z>4.5$ \citep[e.g.][]{richards2002} further
indicating
that the quasars selected in this study are similar to those
selected
in the past.  Both
NDWFS J142516.3+325409 and NDWFS J142937.9+330416 lie quite close
to the
stellar locus in these colors, however, illustrating one of the main
difficulties of
traditional optical color selection techniques.

\begin{figure}[ht]
\centering{\includegraphics[angle=0, width=3.5in]{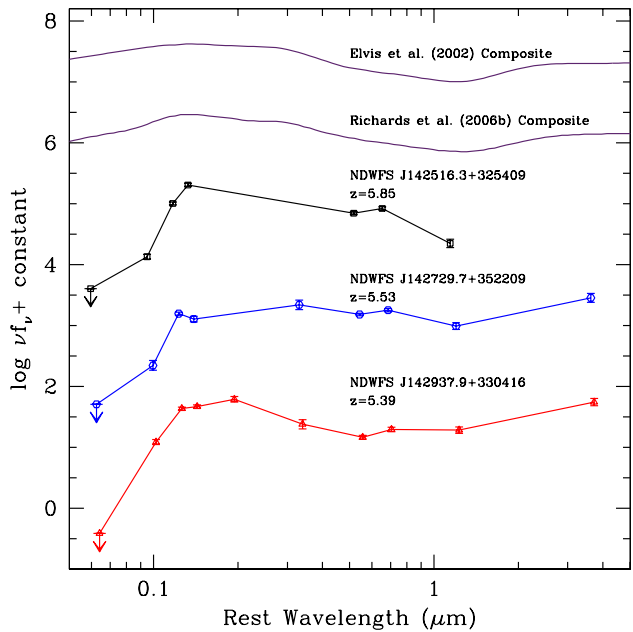}}
\caption{ \scriptsize  Broad-band spectral energy distributions of
the three high-redshift quasars presented here.  The quasars are
ordered by redshift with the highest-redshift object on the top.
The top two lines show two composite quasar spectral energy distributions for comparison; the top line shows the radio-quie
t quasar SED from \citet{elvis2002} while the bottom comparison shows the composite of {\it Spitzer} observed SDSS quasars 
from \citet{richards2006b}.
The high-redshift SEDs are very similar to each other and the
low-redshift composites are broadly similar to the SEDs of the high-redshift objects although the SEDs of the three $z>5$ o
bjects presented here have much stronger breaks in the blue from absorption from neutral hydrogen. The large ratio between 
the 4.5\micron
and 8.0\micron flux of the highest-redshift object, NDWFS J142516.3+325409 is likely a result of
low signal-to-noise ratio photometry as this object is close to the detection limit in that band.  }
\label{fig:sed}
\end{figure}

Figure \ref{fig:sed} shows the spectral energy distribution from
the observed $B_W$-band to 24 \micron (approximately rest-frame 600
$\mbox{\AA}$ - 3.7 \micron) for each of the high-redshift
quasars identified in this
work.  Again, all three of these objects were selected based upon their mid-infrared colors;  the two lowest-redshift quasars were also selected based on their 24\micron fluxes.  For comparison, we have also plotted the
average quasar SEDs from two studies in the literature; the top line shows the \citet{elvis2002} radio-quiet 
quasar composite while the lower line shows the average SED of {\it Spitzer}-observed SDSS
quasars from \citet{richards2006b}.  
The broad photometric
properties of the high-redshift quasars are quite similar to each
other as well as to the low-redshift composite spectra.  The main
difference between the low- and high-redshift objects is the lack of
absorption due to neutral hydrogen at the highest frequencies in the
composite spectra. The highest redshift object, NDWFS J142516.3+325409, shows a strong break between observed 4.5\micron and 8.0\micron, but this is likely 
due to poor signal-to-noise ratio in the 8.0\micron photometry as this object is near the flux limit
of the 8.0\micron imaging.   The rest-frame UV to optical color of NDWFS J142729.7+352209 is redder than the other two quasars studied here as well as both of the comparison composites, possibly indicating enhanced dust extinction in this object.  Compared to a sample of 58 quasars at $3<z<3.5$ from AGES, the rest-frame UV to optical color of NDWFS J142729.7+352209 is not unique;  it falls well within the distribution of colors of lower-redshift objects. 

\begin{deluxetable}{cccc}
\tablecolumns{4}
\tablenum{3}
\tablewidth{0pt}
\tabletypesize{\scriptsize}
\tablecaption{Luminosities of Discovered Quasars}
\tablehead{
  \colhead{Object Name} &
  \colhead{Redshift} &
\colhead{$M_B$} &
\colhead{$M_{1450}$}}
\startdata 
NDWFS J142937.9+330416 &  5.39 &       -26.00 &       -25.52 \\ 
NDWFS J142729.7+352209 &  5.53 &       -25.15 &       -24.67 \\ 
NDWFS J142516.3+325409 &  5.85 &       -26.52 &       -26.03
\enddata
\end{deluxetable}

\begin{figure}[ht]
\centering{\includegraphics[angle=0, width=3.5in]{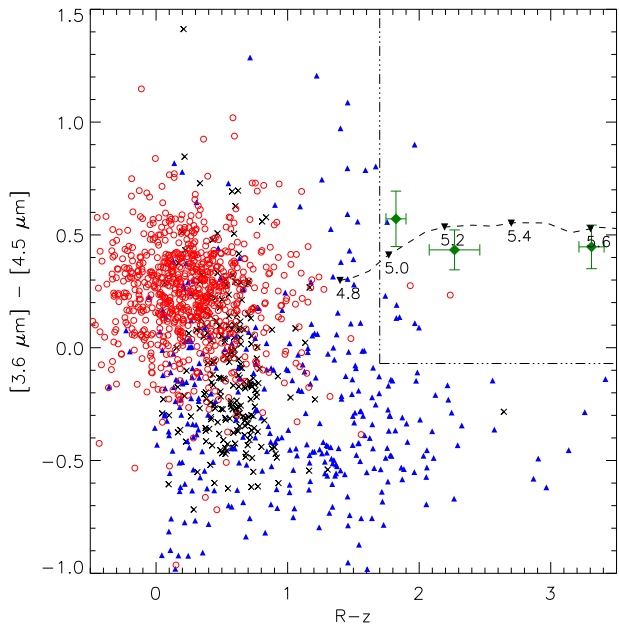}}
\caption{ \scriptsize  Optical versus mid-infrared color-color space
of AGES spectroscopically observed point sources. The symbols
and colors
are as in Figure \ref{fig:riiz}. The dashed line
illustrates the color
track of quasars above $z\sim4.8$. The $z>5$ quasars
are well-separated from the loci of low-redshift galaxies, stars,
and lower-redshift quasars.  The region defined by the dot-dashed line
illustrates one way to select
high-redshift quasars by combining optical and mid-infrared
photometry. The minimum efficiency for selecting $z>5$ quasars in this region is 17\% with the largest  portion of the cont
amination 
arising due to stars with low signal-to-noise IRAC photometry which scatter to redder $[3.6]-[4.5]$ colors. 
Deeper mid-infrared imaging would increase the color measurement accuracy, significantly reducing the stellar contamination
 and increasing the 
quasar selection efficiency. }
\label{fig:rzirac}
\end{figure}

Figure \ref{fig:rzirac} shows the optical versus mid-infrared color-color
space for AGES point sources.  In this color space, all three of
the discovered $z>5$ quasars are separated from the locus of
low-redshift galaxies, Galactic stars, and $z<5$ quasars.
We show one
possible selection method for high-redshift quasars by the dot-dashed line in
Figure \ref{fig:rzirac}.  Within this region, the AGES source
catalog contains
19 point sources with $I<22$ mag;  14 of these were spectroscopically observed.
Of these 14 targets, 9 objects are stars, 2 are lower-redshift
($z=2.11, 3.57$) quasars, and 3 have $z>5$.  Assuming
the 4 objects without redshifts are not located at high redshift,
this selection results in a minimum efficiency of $\sim17$\%, though with 
only 3 quasars, this efficiency measurement is rather uncertain.    The large contamination from 
stars arises due to increasing photometric errors in the $[3.6]-[4.5]$ colors as the stars
approach the IRAC flux limit and thus much of the contamination could be mitigated with
a deeper IRAC imaging survey. 

\section{Discussion}

The quasars reported here add three new low-luminosity, high-redshift
quasars to the
slowly growing catalog of these objects.  All three of these
objects are
fainter than quasars at similar redshifts found in the SDSS, and
NDWFS J142516.3+325409 is the lowest luminosity $z>5.8$
quasar currently known.  The number of these objects should grow
quickly as
the next generation of deep wide-area surveys are completed,
opening the
door to understanding the nature of low-luminosity quasars near the
epoch of reionization.

As the number of low-luminosity $z>5$ quasars is still small, the details of
the luminosity function of these objects is poorly constrained.
Extrapolating the \citet{fan2001IV}  quasar luminosity
function (QLF), determined using $3.5<z<5$ quasars with 
$-27.5<\mbox{M}_{\mbox{\scriptsize 1450}}<-25.5$,  to higher redshifts
and lower luminosities, 
we would expect a density of 0.28 deg$^{-2}$ to the optical
flux limit of the AGES spectroscopy or 2.2 quasars with $z>5$ in the
AGES survey area for a complete, optically limited, survey.

In order to estimate the effect of the IRAC
flux limits on the expected number of quasars in our survey,
we create a sample of 58 quasars at $3<z<3.5$ from the full AGES catalog
with existing FLAMEX photometry.  These objects include quasars selected using the
full suite of AGES selection techniques including X-ray, radio, or 24\micron fluxes or their
optical or mid-infrared colors and thus likely exhibit a broad range of broad-band photometric
properties.  We augment this sample with 17 quasars in
the same redshift range listed in the SDSS Third Quasar catalog
\citep{schneider2005} which were also detected in 2MASS. 
For each object in this sample, we convert the observed $R-K_s$
(or $r-K_s$ for SDSS quasars) to a rest-frame ultraviolet-to-optical
spectral slope, $\alpha_{UO}$, by assuming the measured broad-band colors 
are the result of a pure power-law SED with  $f_\nu \propto \nu^\alpha$.  
Next, we create a mock catalog of high-redshift quasars using the
\citet{fan2001IV} QLF to assign each mock object a redshift and
UV luminosity at 1450 $\mbox{\AA}$.  The observed distribution
of UV-to-optical spectral slopes defines the distribution of $k$-corrections that 
are applied to the mock catalog to convert the rest-frame luminosity at 1450\AA\,
to the observed flux at 3.6\micron for each object.  We then use
the observed distribution of mid-infrared spectral indices,
$\alpha_{MIR}=0.73\pm0.84$, measured by \citet{stern2005},
to assign each mock quasar a flux at 4.5, 5.8, and 8.0 \micron.
Finally, we apply the AGES mid-infrared flux limits and selection criteria
to measure the fraction of high-redshift objects which are missed
by the AGES mid-infrared selection criteria.  We find that approximately 35\%
of the mock quasars that pass the $I<22$ mag optical flux limits 
are missed when the IRAC flux limits are included.  It may be a concern that the addition
of the SDSS quasars, which, unlike the AGES sample, were all selected based on their optical  
colors, may bias this estimate; we have verified that omitting these objects from the 
calculation does not affect the final completeness more than a few percent.  It should be 
noted that the above calculation assumes quasars at $z>5$ have the same broad-band photometric 
properties as $z\sim3$ objects; this assumption can only be tested after a large sample of 
high-redshift quasars with multi-wavelength photometry is collected.  In the $I<22$ AGES 
sample of 2153 mid-infrared selected objects, 83\% of the 
targets obtained valid redshifts.  Combining this 83\% spectroscopic completeness with 
the effects of the mid-infrared flux limits, we estimate that our overall completeness is  
54\% and thus we would expect to find only 1.2 quasars at $z>5$ for a QLF with a slope of
$\Psi \propto L^{-2.6}$.  To then find three quasars is somewhat unlikely, since the Poisson
probability of finding 3 or more objects when 1.2 are expected is only 12\%.

Based on a sample of 12  quasars at $z>5.8$, 
\citet{xfanIII} found that the high-redshift QLF was best fit by a
steeper QLF ($\Psi \propto L^{-3.2}$) than that 
measured by \citet{fan2001IV}.
If we instead use the best fit bright-end slope and normalization
from \citet{xfanIII} but keep the redshift evolution of the quasar
number density determined by \citet{fan2001IV}, we would expect
to find 6.2 quasars at $z>5$ in a complete optically-limited 
survey of the AGES area.
This value drops to 3.5 when the effects of our mid-infrared flux
limits and spectroscopic incompleteness are added. This value agrees quite
well with the 3 objects found in our survey, but, at this point, this agreement is only suggestive due 
to the small number of quasars in our sample and the small sample size  and luminosity range used by \citet{xfanIII} to determine this QLF slope.  \citet{richards2006} found the QLF slope to
evolve from $\beta=-3.1$ (where $\Psi \propto L^\beta$) at $z<2.4$ to progressively flatter slopes at higher redshift such that at
$z\sim5$, one would predict $\beta \gtorder -2.4$.  If the QLF at $5<z<6$ is, indeed, as steep at $L^{-3}$, the trend measured by \citet{richards2006} must break down beyond $z\sim4.5$, but a larger sample
of low-luminosity objects is required to place any robust constraints on the evolution 
of the QLF to $z\sim5$.

\begin{figure}[ht]
\centering{\includegraphics[angle=0, width=3.5in]{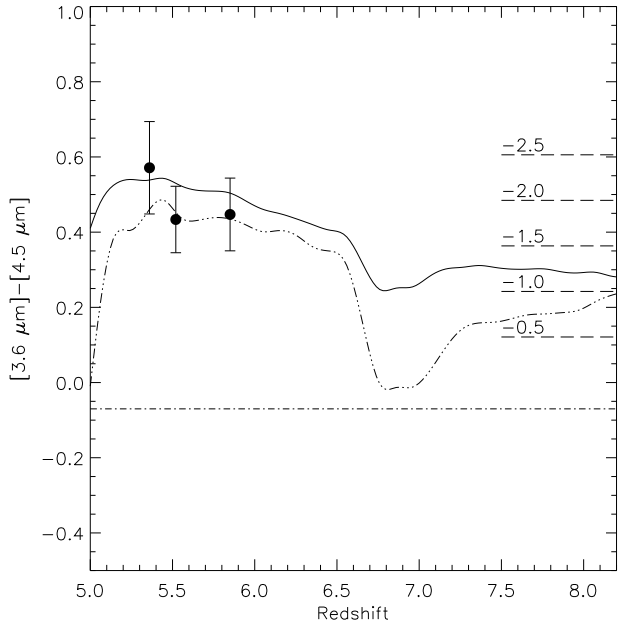}}
\caption{ \scriptsize $[3.6]-[4.5]$ color of quasars to
very high redshift based on the
SDSS quasar composite \citep[solid line,][]{vandenburk2001} or a $\alpha=-0.5$ power-law combined with the Balmer emission 
lines observed in the SDSS composite spectrum (dash-dot-dot-dot line).  The color cut imposed
by the mid-infrared selection of quasars in AGES
is shown by the dot-dashed line.   Longward
of rest-frame $5000\mbox{\AA}$, the SDSS composite is dominated by low-luminosity AGNs  and is likely
heavily contaminated by host-galaxy light, making the colors slightly redder. 
For comparison, the location of
the three $z>5$ quasars discovered here are marked and the dotted lines on the right mark the 
colors of pure power-law spectra with different values of $\alpha_\nu$.
The rapid rise at $z=5$ occurs as the H$\alpha$ emission line passes from the 3.6\micron band into the 4.5\micron bandpass.
  Similarly,  as the H$\alpha$ emission line redshifts
out of the 4.5\micron bandpass near $z\sim6.5$, the colors drop quickly; the colors become redder again near 
$z\sim7$ as the H$\beta$ enters the 4.5\micron bandpass.
Even at very high redshifts, the colors
of quasars predicted by both tracks fall within the region used to mid-infrared select
quasars in our survey.}
\label{fig:hizcolor}
\end{figure}

We have shown that mid-infrared selection of quasars can be useful
in the search for high-redshift objects to $z\sim6$, thus avoiding
some of the key problems inherent in optically based high-redshift
quasar searches.  At $z\gtorder7$, optical selection of quasars is
impossible as the Ly$\alpha$ emission from objects at these redshifts
is shifted out of the $z'$-band.  In the near-infrared, the existing
large area surveys (primarily 2MASS) are too shallow to detect even the
brightest high-redshift quasars found by the SDSS. In the near future, 
the UKIDSS survey will image 4000 deg$^2$ of the sky 
to 3 mag deeper than 2MASS and
should find about 10 quasars  with $5.8<z<7.2$ \citep{warren2002}.
While near-infrared spectroscopy is ultimately the only means
by which one can ensure any candidate is located at high redshift,
the use of mid-infrared colors in the selection of these objects
may provide a valuable tool in separating stars
from quasars at the highest redshifts.  Figure \ref{fig:hizcolor}
illustrates the expected  $[3.6]-[4.5]$ color of 
quasars to $z\sim8$
based upon the locally determined SDSS quasar composite spectrum
\citep{vandenburk2001}.  We also show the colors predicted from a power-law ($\alpha_\nu = -0.5$)
continuum with Balmer emission lines having the same strength and width of those measured on the 
SDSS composite spectrum. 

As the strong emission from the H$\alpha$ emission line redshifts into the 
4.5\micron band, the colors of quasars redden quickly near $z\sim5$.  
Correspondingly, the sudden drop
in color at $z\sim6.5$ occurs as the strong H$\alpha$ emission
line redshifts out of the 4.5\micron band; 
by $z\sim7$, the colors of high-redshift quasars
redden as H$\beta$ redshifts from the 3.6\micron band 
into the 4.5\micron IRAC bandpass.  The overall trend toward bluer colors exhibited by the SDSS
composite 
across this redshift range occurs as the IRAC passbands probe the transition between 
a spectral slope of $\alpha_\nu = -0.5$ measured blueward of 5000\AA\, and of $\alpha_\nu=-2.5$ 
redward of this wavelength in 
the SDSS composite spectrum.  This transition likely occurs due to contamination of the 
AGN light by the host-galaxy in the low-redshift AGNs used to construct the composite spectrum.  
This contamination makes the colors predicted using the \citet{vandenburk2001} composite 
generally redder than those expected using the  simple model 
consisting only of a power-law continuum and
Balmer emission lines.  For both tracks, the mid-infrared colors of very high-redshift quasars
remain redder than stars ($[3.6]-[4.5]\sim-0.5$) and low-redshift galaxies throughout the 
redshift range shown. As
the large number of deep, wide-area, {\it Spitzer} fields with
corresponding deep optical and near-infrared photometry become available, the
ability to build large samples of high-redshift quasars,
including low-luminosity objects, will allow us, for the first time,
to probe the statistics of low-luminosity quasars at high redshift.

\acknowledgments

We greatly appreciate the help of Mark Dickinson and Steve Dawson with the Keck observations used here.  RJC is funded through a National Science Foundation Graduate
Research Fellowship.  XF acknowledges support from NSF grant AST
03-07384 and a Packard Fellowship for Science and 
Engineering.   Both DJE and XF receive support from an  Alfred P. Sloan Research Fellowship.  AHG acknowledges support from an NSF Small Grant for 
Exploratory Research under award AST-0436681.

Observations reported here were obtained at the MMT Observatory,
a joint facility of the Smithsonian Institution and the University
of Arizona.  This work made use of images and/or data products
provided by the NOAO Deep Wide-Field Survey \citep{jannuzidey1999},
which is supported by the National Optical Astronomy Observatory
(NOAO). NOAO is operated by AURA, Inc., under a cooperative agreement
with the National Science Foundation.  This work is based in part on observations 
made with the {\it Spitzer Space Telescope,} which is operated by the Jet Propulsion 
Laboratory, California Institute of Technology under a contract with NASA. Support 
for this work was provided by NASA through an award issued by JPL/Caltech.
Some of the data presented here were
obtained at the W.M. Keck Observatory, which is operated as a
scientific partnership among the California Institute of Technology,
the University of California, and the National Aeronautics and
Space Administration.  The Observatory was
made possible by the generous financial support of the W.M. Keck
Foundation.  The authors wish to recognize and acknowledge the very significant cultural role 
and reverence that the summit of Mauna  Kea has always had within the indigenous Hawaiian 
community.  We are most fortunate to have the opportunity to conduct observations from this
mountain.

\clearpage


\begin{thebibliography}{}

\bibitem[Abazajian et al.(2003)]{a2003}  Abazajian, K., et al.\
2003, \aj, 126, 2081
\bibitem[Abazajian et al.(2004a)]{a2004a}  Abazajian, K., et al.\
2004, \aj, 128, 502
\bibitem[Abazajian et al.(2005)]{a2004b} Abazajian, K., et al.\
2005, \aj, 129, 1755
\bibitem[Adelman-McCarthy et al.(2005)]{a2005} Adelman-McCarthy,
J.~K.\ 2005, submitted, arXiv:astro-ph/0507711
\bibitem[Barger et al.(2002)]{barger2002}  Barger, A.~J., Cowie,
L.~L., Brandt, W.~N., Capak, P., Garmire, G.~P., Hornschemeier,
A.~E., Steffen, A.~T., \& Wehner, E.~H.\ 2002, \aj, 124, 1839
\bibitem[Barger et al.(2003)]{barger2003}  Barger, A.~J., Cowie,
L.~L., Capak, P., Alexander, D.~M., Bauer, F.~E., Brandt, W.~N.,
Garmire, G.~P., \& Hornschemeier, A.~E.\ 2003, \apjl, 584, L61
\bibitem[Barger et al.(2005)]{barger2005} Barger, A.~J., Cowie,
L.~L., Mushotzky, R.~F., Yang, Y., Wang, W.-H., Steffen, A.~T., \&
Capak, P.\ 2005, \aj, 129, 578
\bibitem[Bertin \& Arnouts(1996)]{sextractor} Bertin, E., \& 
Arnouts, S.\ 1996, \aaps, 117, 393
\bibitem[Boyle et al.(2000)]{boyle2000} Boyle, B.~J., Shanks, T., 
Croom, S.~M., Smith, R.~J., Miller, L., Loaring, N., \& Heymans, C.\ 2000, 
\mnras, 317, 1014 
\bibitem[Brown et al.(2006)]{brown2006} Brown, M.~J.~I., et al.\
2006, in press, arXiv:astro-ph/0510504
\bibitem[Chiu et al.(2005)]{chiu2005} Chiu, K., et al.\ 2005, 
\aj, 130, 13 
\bibitem[Cristiani et al.(2004)]{cristiani2004} Cristiani, S.,
et al.\ 2004, \apjl, 600, L119
\bibitem[Croom et al.(2004)]{croom2004} Croom, S.~M., Smith, R.~J.,
Boyle, B.~J., Shanks, T., Miller, L., Outram, P.~J., \& Loaring,
N.~S.\ 2004, \mnras, 349, 1397
\bibitem[Djorgovski et al.(2003)]{djorgovski2003} Djorgovski, S.~G.,
Stern, D., Mahabal, A.~A., \& Brunner, R.\ 2003, \apj, 596, 67
\bibitem[Eisenhardt et al.(2004)]{eisenhardt2004} Eisenhardt, P.~R.,
et al.\ 2004, \apjs, 154, 48
\bibitem[Elston et al.(2005)]{elston2005} Elston, R.~J., et al. 2005,
in press, arXiv:astro-ph/0511249 
\bibitem[Elvis, Risaliti, \& Zamorani(2002)]{elvis2002}  Elvis, M.,               
Risaliti, G., \&   Zamorani, G. 2002, \apjl, 565, L75   
\bibitem[Faber et al.(2003)]{faber2003} Faber, S.~M., et al.\ 2003,
\procspie, 4841, 1657
\bibitem[Fabricant et al.(1998)]{fabricant1998} Fabricant, D.~G.,
Hertz, E.~N., Szentgyorgyi, A.~H., Fata, R.~G., Roll, J.~B., \&
Zajac, J.~M.\ 1998, \procspie, 3355, 285
\bibitem[Fabricant et al.(2005)]{fabricant2005} Fabricant, D.,
et al.\ 2005, \pasp, 117, 1411
\bibitem[Fan et al.(2000)]{fan2000} Fan, X., et al.\ 2000, \aj,
120, 1167
\bibitem[Fan et al.(2001a)]{fan2001III} Fan, X., et al.\ 2001a,
\aj, 121, 31
\bibitem[Fan et al.(2001b)]{fan2001IV}  Fan, X., et al.\ 2001b, \aj,
121, 54
\bibitem[Fan et al.(2001c)]{xfanI} Fan, X., et al.\ 2001c, \aj,
122, 2833
\bibitem[Fan et al.(2003)]{xfanII} Fan, X., et al.\ 2003, \aj,
125, 1649
\bibitem[Fan et al.(2004)]{xfanIII} Fan, X., et al.\ 2004, \aj,
128, 515
\bibitem[Fan et al.(2005)]{xfanIV} Fan, X., et al.\ 2005, in press,
arXiv:astro-ph/0512080
\bibitem[Fazio et al.(2004)]{fazio2004}  Fazio, G.~G., et al.\
2004, \apjs, 154, 10
\bibitem[Houck et al.(2005)]{houck2005} Houck, J.~R., et al.\ 2005,
\apjl, 622, L105
\bibitem[Ivezi{\' c} et al.(2004)]{ivezic2004} Ivezi{\' c}, {\v Z}.,
et al.\ 2004, Astronomische Nachrichten, 325, 583
\bibitem[Jannuzi \& Dey(1999)]{jannuzidey1999} Jannuzi, B.~T., \&
Dey, A.\ 1999, ASP Conf.~Ser.~191: Photometric Redshifts and the
Detection of High Redshift Galaxies, 191, 111
\bibitem[Jiang et al.(2006)]{jiang2005} Jiang, L., et al.\ 2006, 
ArXiv Astrophysics e-prints, arXiv:astro-ph/0602569
\bibitem[Kron(1980)]{kron1980} Kron, R.~G.\ 1980, \apjs, 43, 305 
\bibitem[Kurucz(1993)]{kurucz} Kurucz, R.\ 1993, ATLAS9 Stellar
Atmosphere Programs and 2 km/s grid.~Kurucz CD-ROM No.~13.~
Cambridge, Mass.: Smithsonian Astrophysical Observatory, 1993, 13,
\bibitem[Mahabal et al.(2005)]{mahabal2005} Mahabal, A., Stern, D.,
Bogosavljevi{\'c}, M., Djorgovski, S.~G., \& Thompson, D.\ 2005,
\apjl, 634, L9
\bibitem[Murray et al.(2005)]{murray2005} Murray, S.~S., et al.\
2005, \apjs, 161, 1
\bibitem[Oke (1974)]{oke1974} Oke, J.~B. 1974, \apjs, 27, 21
\bibitem[Richards et al.(2002)]{richards2002} Richards, G.~T.,
et al.\ 2002, \aj, 123, 2945
\bibitem[Richards et al.(2005)]{richards2005} Richards, G.~T.,
et al.\ 2005, \mnras, 360, 839
\bibitem[Richards et al.(2006)]{richards2006} Richards, G.~T., et 
al.\ 2006, ArXiv Astrophysics e-prints, arXiv:astro-ph/0601434 
\bibitem[Richards et al.(2006b)]{richards2006b} Richards, G.~T., et 
al.\ 2006b, ArXiv Astrophysics e-prints, arXiv:astro-ph/0601558 
\bibitem[Rieke et al.(2004)]{rieke2004} Rieke, G.~H., et al.\ 2004,
\apjs, 154, 25
\bibitem[Roll et al.(1998)]{roll1998} Roll, J.~B., Fabricant, D.~G.,
\& McLeod, B.~A.\ 1998, \procspie, 3355, 324
\bibitem[Schlegel, Finkbeiner, \& Davis (1998)]{sfd1998} Schlegel, D.~J., Finkbeiner,
D.~P., \& Davis, M.\ 1998, \apj, 500, 525
\bibitem[Schmidt \& Green(1983)]{pgbqs} Schmidt, M., \& 
Green, R.~F.\ 1983, \apj, 269, 352
\bibitem[Schneider et al.(2005)]{schneider2005} Schneider, D.~P., et 
al.\ 2005, \aj, 130, 367 
\bibitem[Sharp et al.(2004)]{sharp2004} Sharp, R.~G., Crampton,
D., Hook, I.~M., \& McMahon, R.~G.\ 2004, \mnras, 350, 449
\bibitem[Songaila \& Cowie(2002)]{songaila2002} Songaila, A., \&
Cowie, L.~L.\ 2002, \aj, 123, 2183
\bibitem[Stern et al.(2000)]{stern2000} Stern, D., Spinrad, H.,
Eisenhardt, P., Bunker, A.~J., Dawson, S., Stanford, S.~A., \&
Elston, R.\ 2000, \apjl, 533, L75
\bibitem[Stern et al.(2005)]{stern2005} Stern, D., et al.\ 2005,
\apj, 631, 163
\bibitem[Stoughton et al.(2002b)]{stoughton2002b}  Stoughton, C.,
et al.\ 2002, \procspie, 4836, 339
\bibitem[Vanden Berk et al.(2001)]{vandenburk2001} Vanden Berk,
D.~E., et al.\ 2001, \aj, 122, 549
\bibitem[Warren \& Hewett(2002)]{warren2002} Warren, S., \& 
Hewett, P.\ 2002, ASP Conf.~Ser.~283: A New Era in Cosmology, 283, 369 
\bibitem[Williams et al.(2004)]{williams2004} Williams, G.~G.,
Olszewski, E., Lesser, M.~P., \& Burge, J.~H.\ 2004, \procspie,
5492, 787
\bibitem[Willott et al.(2005)]{willott2005}  Willott, C.~J.,
Delfosse, X., Forveille, T., Delorme, P., \& Gwyn, S.~D.~J.\ 2005,
\apj, 633, 630
\bibitem[Wolf et al.(2003)]{wolf2003} Wolf, C., Wisotzki, L., Borch,
A., Dye, S., Kleinheinrich, M., \& Meisenheimer, K.\ 2003, \aap,
408, 499
\bibitem[York et al.(2000)]{york2000} York, D.~G., et al.\ 2000,
\aj, 120, 1579
\bibitem[Zheng et al.(2000)]{zheng2000} Zheng, W., et al.\ 2000, 
\aj, 120, 1607 
 

\end{thebibliography}
\end{document}